\documentclass[11pt]{article}
\usepackage{amsmath,amssymb,color,graphics,epsfig}
\usepackage{amsfonts}
\newcommand{\be}{\begin{equation}}
\newcommand{\ee}{\end{equation}}
\newcommand{\bea}{\setlength\arraycolsep{2pt} \begin{eqnarray}}
\newcommand{\eea}{\end{eqnarray}}
\newcommand{\nn}{\nonumber}

\def\ft#1#2{{\textstyle{\frac{\scriptstyle #1}{\scriptstyle #2} } }}
\def\fft#1#2{{\frac{#1}{#2}}}

\def\0{{\sst{(0)}}}
\def\1{{\sst{(1)}}}
\def\2{{\sst{(2)}}}
\def\3{{\sst{(3)}}}
\def\4{{\sst{(4)}}}
\def\5{{\sst{(5)}}}
\def\6{{\sst{(6)}}}
\def\7{{\sst{(7)}}}
\def\8{{\sst{(8)}}}
\def\sst#1{{\scriptscriptstyle #1}}

\begin{document}

\begin{flushright}
%\hfill{KIAS-P12028}
 %\hfill{
%\bf hep-th/yymmnnn}
\end{flushright}

\vspace{25pt}
\begin{center}
{\large {\bf Criticality in Einstein-Gauss-Bonnet Gravity: Gravity without Graviton }}

\vspace{10pt}
 Zhong-Ying Fan$^{1}$, Bin Chen$^{1,2,3}$ and Hong L\"u$^{4}$

\vspace{10pt}
{\it $^{1}$Center for High Energy Physics, Peking University, No.5 Yiheyuan Rd,\\}
{\it  Beijing 100871, P. R. China\\}
\smallskip
{\it $^{2}$Department of Physics and State Key Laboratory of Nuclear Physics and Technology,\\}
{\it Peking University, No.5 Yiheyuan Rd, Beijing 100871, P.R. China\\}
\smallskip
{\it $^{3}$Collaborative Innovation Center of Quantum Matter, No.5 Yiheyuan Rd,\\}
{\it  Beijing 100871, P. R. China\\}

{\it $^{4}$Center for Advanced Quantum Studies, Department of Physics, \\
Beijing Normal University, Beijing 100875, China}

\vspace{40pt}

\underline{ABSTRACT}
\end{center}

General Einstein-Gauss-Bonnet gravity with a cosmological constant allows two (A)dS spacetimes as its vacuum solutions.  We find a critical point in the parameter space where the two (A)dS spacetimes coalesce into one and the linearized perturbations lack any bilinear kinetic terms.  The vacuum perturbations hence loose their interpretation as linear graviton modes at the critical point. Nevertheless, the critical theory admits black hole solutions due to the nonlinear effect. We also consider Einstein gravity extended with general quadratic curvature invariants and obtain critical points where the theory has no bilinear kinetic terms for either the scalar trace mode or the transverse modes.  Such critical phenomena are expected to occur frequently in general higher derivative gravities.

\vfill {\footnotesize Emails: fanzhy@pku.edu.cn\ \ \ bchen01@pku.edu.cn \ \ \ mrhonglu@gmail.com}

\thispagestyle{empty}

\pagebreak

\tableofcontents
\addtocontents{toc}{\protect\setcounter{tocdepth}{2}}

%%%%%%%%%%%%%%%%%%%%%%%%%%%%%%%%%%%%%%%%

%\newpage
%%%%%%%%%%%%%%%%%%%%%%%%%%%%%%%%%%%%%%%%

%\vspace{2cm}

\section{Introduction}

Einstein gravity with a cosmological constant may be viewed as the simplest dynamical theory of the metric under the  principle of general coordinate invariance.  Owing to the existence of a fundamental constant of length, namely the Planck length $\ell_p=\sqrt{G_{\rm N}}$, where $G_{\rm N}$ is the Newton's constant, it is natural to consider higher-derivative extensions to Einstein gravity.  This should be contrasted with quantum field theory where such a ``minimum'' length scale is absent and hence it is ``unnatural'' to consider higher-derivative terms. String theory predicts that such higher-derivative structures are inevitable in its low-energy effective theory.  The explicit structure however is hard to determine.

Introducing higher-derivative terms to Einstein gravity can have important advantages. It was shown that Einstein gravity extended with quadratic curvature invariants in four dimensions can be renormalizable for appropriate coupling constants \cite{Stelle:1976gc,Stelle:1977ry}; however, the theory contains ghost-like massive spin-2 modes. When a cosmological constant is included, there exists a critical point of the parameter space \cite{Lu:2011zk,Deser:2011xc} of the coupling constants for which the ghost-like massive graviton becomes a logarithmic mode.  In three dimensions, imposing a strong boundary condition to get rid of this mode may yield a consistent quantum theory of gravity, whose degrees of freedom exist only in the boundary of the anti-de Sitter (AdS$_3$) spacetime \cite{Li:2008dq,Maloney:2009ck}. This procedure may not be possible in four or higher dimensions; rather the theories are expected to be dual to some logarithmic conformal field theories on the boundary of AdS. (See {\it e.g.}~reviews \cite{Johansson:2012fs,Grumiller:2013at,Hogervorst:2016itc}.)

     In higher dimensions, there exist further special combinations of higher-order curvature invariants for which the linearized theories involve only two derivatives, and hence ghost excitations can be absent.  These are Gauss-Bonnet or Lovelock gravities \cite{Lovelock:1971yv}.  Einstein-Gauss-Bonnet (EGB) gravity contains two non-trivial parameters, namely the bare cosmological constant $\Lambda_0$ and the coupling constant $\gamma$ of the Gauss-Bonnet term.  In general, there exists two (A)dS vacua in the EGB theory.  In section 2, we consider linearized gravity in these vacua and find that the linear modes have positive kinetic energy in one vacuum whilst they have negative energy in the other.  At some critical point, the two (A)dS coalesce into one and the effective coupling for the kinetic term vanishes.  The theory at the critical point thus does not have propagators and hence the linear modes cannot be viewed as gravitons.  We then derive the perturbative equations of motion at the quadratic order.  Furthermore we  obtain the most general static solutions with spherical/toric/hyperbolic topologies.  Using the Wald formalism \cite{wald1,wald2}, we find that these solutions have no negative mass, indicating that the theory may not have nonlinear ghost modes.  One of the solutions describes a black hole, which was obtained in \cite{Crisostomo:2000bb}, where general Lovelock gravities that addmit a single (A)dS vacuum were classified and studied. We analyse its global structure and the thermodynamical phase transition.

   In section 3, we consider Einstein gravity extended with general quadratic curvature invariants, with additional $\alpha R^2 + \beta R^{\mu\nu} R_{\mu\nu}$ terms.  We find that there also exists a critical point where the linearized equation of motion for the scalar trace mode is automatically satisfied.  In this case the theory does not have a kinetic term for the scalar mode.  We also obtain another critical point where the linearized equations of motion of the theory involve only the trace scalar mode, whilst there is no kinetic term for any transverse mode.  Note that the critical case where all linear perturbations have no kinetic terms can only occur when $\alpha=0=\beta$, in other words, in the EGB theory. We then consider the deviations from the critical points in section 4, which helps us to understand the integration constants of the solutions at the critical point in the more general setting.  We conclude the paper in section 5.

\section{Einstein-Gauss-Bonnet gravity}

In this section, we consider Einstein-Gauss-Bonnet (EGB) gravity without  matter fields. The Lagrangian density is given by
\be
\mathcal{L}=\kappa \sqrt{-g} \Big[\big(R-2\Lambda_0 \big)+\gamma\big(R^2-4R_{\mu\nu}R^{\mu\nu}+R_{\mu\nu\rho\sigma}R^{\mu\nu\rho\sigma} \big)\Big] \,,
\ee
where $\kappa=1/(16\pi\, G_{\rm N}) >0 $ with $G_{\rm N}$ being the Newton's constant, and $\Lambda_0$ is the bare cosmological constant. The parameter $\gamma$ has a dimension of length squared, and in string theory, it is related to the string tension or  the coupling constant $\alpha'$ in string worldsheet action.

The equations of motion are $E_{\mu\nu} = 0$  and
\bea
E_{\mu\nu} &\equiv &G_{\mu\nu} + \Lambda_0 g_{\mu\nu}  +
2\gamma \big(R R_{\mu\nu} - 2 R_{\mu\sigma\nu\rho} R^{\sigma\rho} +
R_{\mu\sigma\rho\lambda} R_\nu{}^{\sigma\rho\lambda} - R_{\mu\rho} R_{\nu}{}^\rho\big)\cr
&& - \ft12\gamma g_{\mu\nu} \big(R^2-4R_{\mu\nu}R^{\mu\nu}+R_{\mu\nu\rho\sigma}R^{\mu\nu\rho\sigma} \big)\,,
\eea
where $G_{\mu\nu}=R_{\mu\nu} - \fft12 g_{\mu\nu} R$ is the Einstein tensor.  It is well known that there exist two distinct (A)dS vacua for generic values of the coupling constants. The effective cosmological constant $\Lambda$ satisfies a quadratic algebraic equation
\be
\kappa\big[\ft 12\big(\Lambda-\Lambda_0 \big)+\Delta_0\,\Lambda^2\big]=0\,,\qquad \hbox{with}\qquad \Delta_0\equiv \fft{(D-3)(D-4)}{(D-1)(D-2)}\gamma \,.\label{vacua}
\ee
Thus the two effective cosmological constants are given by
\be
\Lambda_\pm =\fft{\pm\sqrt{1 + 8\Delta_0\,\Lambda_0} - 1}{4\Delta_0}\,.\label{deltapm}
\ee
When $\gamma=0$ and hence $\Lambda_0=0$, one of the (A)dS spacetimes becomes Minkowski.  When $\Lambda_0=-1/(8\Delta_0)$, the two effective cosmological constants $\Lambda_\pm$ become the same, and the two (A)dS vacua degenerate into one, with the effective cosmological constant
\be
\Lambda_+=\Lambda_-=\Lambda^* \equiv 2\Lambda_0\,.\label{2in1}
\ee
Note that the reality condition for $\Lambda_\pm$ requires that $8\Delta_0\Lambda_0\ge -1$.  When this condition is not satisfied, the theory then does not admit any maximally-symmetric spacetime as its vacuum solution \cite{Canfora:2013xsa}.

General Lovelock gravities with only a single (A)dS vacuum were classified and studied in
\cite{Crisostomo:2000bb}.

\subsection{Linearized gravity}

We now study the linearized equations of motion of the metric perturbation
\be
g_{\mu\nu}=\bar{g}_{\mu\nu}+h_{\mu\nu} \label{pert}
\ee
around one of the (A)dS vacua for general parameters.  They are simply
\be
\kappa_{\rm eff}\, {\cal G}^L_{\mu\nu}=0\,, \qquad \kappa_{\rm eff}=\kappa(1+4\Delta_0 \Lambda)\,.\label{linear1}
\ee
The linearized Einstein tensor around the (A)dS vacuum is given by
\be
{\cal G}^L_{\mu\nu}=R^L_{\mu\nu}-\ft 12 \bar{g}_{\mu\nu}R^L-
\fft{2\Lambda}{D-2}h_{\mu\nu} \,,\label{gmunuL}
\ee
where the linearized Ricci tensor $R^L_{\mu\nu}$ and scalar curvature $R^L\equiv \big(g^{\mu\nu}R_{\mu\nu} \big)^L$ are respectively
\be R^L_{\mu\nu}=\fft 12 \Big(\bar{\nabla}^{\sigma}\bar{\nabla}_{\mu}h_{\nu\sigma}+
\bar{\nabla}^{\sigma}\bar{\nabla}_{\nu}h_{\mu\sigma}-\bar{\square}h_{\mu\nu}
-\bar{\nabla}_{\mu}\bar{\nabla}_{\nu}h   \Big)\,,\quad R^L=-\bar{\square}h+\bar{\nabla}^\sigma\bar{\nabla}^\mu h_{\mu\sigma}-\fft{2\Lambda}{D-2}h\,.
\label{RL1}
\ee
For later purposes we also give the linearized Riemann tensors
\bea
R^{L\quad\rho}_{\mu\lambda\nu}&=&\bar{\nabla}_{[\lambda}\bar{\nabla}_{\mu]}h^{\rho}_{\nu} +\bar{\nabla}_{[\lambda}\bar{\nabla}_{|\nu|} h^{\rho}_{\mu]}-\bar{\nabla}_{[\lambda}\bar{\nabla}^{\rho}h_{\mu]\nu}\cr
&=& \fft 12\Big[\bar{\nabla}_{\lambda}\bar{\nabla}_{\mu}h^{\rho}_{\nu}+ \bar{\nabla}_{\lambda}\bar{\nabla}_{\nu}h^\rho_{\mu} +\bar{\nabla}_{\mu}\bar{\nabla}^{\rho}h_{\lambda\nu} -\bar{\nabla}_{\mu}\bar{\nabla}_{\lambda}h^\rho_{\nu} -\bar{\nabla}_{\mu}\bar{\nabla}_{\nu}h^\rho_{\lambda} -\bar{\nabla}_{\lambda}\bar{\nabla}^{\rho}h_{\mu\nu} \Big]\,,\cr
R^L_{\mu\lambda\nu\rho}&\equiv& \big( g_{\rho\sigma}R_{\mu\lambda\nu}^{\quad\,\,\, \sigma} \big)^L =\bar{g}_{\rho\sigma}R^{L\quad\sigma}_{\mu\lambda\nu}+
\fft{2\Lambda}{(D-1)(D-2)}\big(\bar{g}_{\mu\nu}h_{\lambda\rho}-\bar{g}_{\nu\lambda}h_{\mu\rho} \big)\cr
 &=&\Big(\bar{\nabla}_{[\lambda}\bar{\nabla}_{|\nu|}h_{\mu]\rho} -\bar{\nabla}_{[\lambda}\bar{\nabla}_{|\rho|}h_{\mu]\nu}\Big) +\fft{2\Lambda}{(D-1)(D-2)} \Big(\bar{g}_{\nu[\mu}h_{\lambda]\rho}-\bar{g}_{\rho[\mu}h_{\lambda]\nu} \Big)\,.
\label{RL2}
\eea
Taking the trace of (\ref{linear1}) gives
\be
\kappa_{\rm eff}\, R^L=0 \,.\label{traceless}
\ee
For the (A)dS background, it is advantageous to take the following gauge choice \cite{Li:2008dq}
\be
\bar{\nabla}^\mu h_{\mu\nu}=\bar{\nabla}_\nu h\,.\label{gaugechoice}
\ee
It follows that $R^L=-\fft{2\Lambda}{D-2}h$ and
\bea
R^L_{\mu\nu}=\Big[-\fft 12\bar{\square}h_{\mu\nu}+\fft{2D\Lambda}{(D-1)(D-2)}h_{\mu\nu}\Big ]+\Big[\fft 12 \bar{\nabla}_\mu\bar{\nabla}_\nu h
-\fft{2\Lambda}{(D-1)(D-2)}\bar{g}_{\mu\nu}h\Big]\,,\nn\\
\mathcal{G}^L_{\mu\nu}=\Big[-\fft 12\bar{\square}h_{\mu\nu}+\fft{2\Lambda}{(D-1)(D-2)}h_{\mu\nu}\Big]+\Big[\fft 12 \bar{\nabla}_\mu\bar{\nabla}_\nu h
+\fft{(D-3)\Lambda}{(D-1)(D-2)}\bar{g}_{\mu\nu}h\Big]\label{RG}\,.
\eea
For generic parameters with $\kappa_{\rm eff}\neq 0$, the trace equation (\ref{traceless}) implies the traceless condition $h=0$ and hence the graviton mode is also transverse.
The linearized equation of motion becomes simply
\be \Big(\bar{\square}-\fft{4\Lambda}{(D-1)(D-2)}\Big)h_{\mu\nu}=0 \,.\label{massless1}\ee
This is the equation of motion satisfied by the massless graviton, in each of the two (A)dS vacua. It is worth pointing out however that the effective coupling constant $\kappa_{\rm eff}$ on the two vacua has opposite signs, namely
\be
\kappa_{\rm eff} = \pm \kappa \sqrt{1 + 8\Delta_0\Lambda_0}\,.
\ee
Thus the linear graviton on the $\Lambda_+$ vacuum has the positive kinetic energy, whilst the one on the $\Lambda_-$ vacuum has the negative kinetic energy, and hence is ghost-like.

In string theory the bare cosmological constant $\Lambda_0$ vanishes, and hence $\Lambda_+=0$ and $\Lambda_-=-1/(2\Delta_0)$.  It follows that the Minkowski vacuum remains ghost free under the $\alpha'$ correction.  Note that $\gamma$ is positive in string theory and hence the other vacuum is AdS, with ghost-like graviton modes. For a non-vanishing $\Lambda_0$, we have $\Lambda_+>\Lambda_-$ for $\gamma>0$ and  $\Lambda_+<\Lambda_-$ for $\gamma<0$.

\subsection{Critical point}

When the parameters satisfy (\ref{2in1}), {\it i.e.}~the Gauss-Bonnet coupling constant $\gamma$ and the bare cosmological constant are related as follows
\be
\Delta_0=-\fft{1}{8\Lambda_0}=-\fft{1}{4\Lambda^*}\,,\label{gammalambda}
\ee
the two (A)dS vacua coalesce into one, with the effective cosmological constant being $2\Lambda_0$. In this case, we have $\kappa_{\rm eff}=0$ and hence the linearized equations of motion in the above subsection are automatically satisfied. The absence of the kinetic term  for the fluctuation $h_{\mu\nu}$ at the quadratic order implies that the theory does not have any propagator, and hence it is no longer proper to take $h_{\mu\nu}$ as the usual graviton modes. We have thus a theory of gravity without graviton.

     As we run the parameter $8\Delta_0\Lambda_0+1\rightarrow 0^+$, we have $\kappa_{\rm eff}\rightarrow 0$. One might expect that $\kappa_{\rm eff}$ becomes negative as one let $8\Delta_0 \Lambda_0+1$ be negative such that the theory has ghost-like mode. However this never happens. Instead, as $8\Delta_0 \Lambda_0+1$ becomes negative, the vacuum spacetime is no longer maximally-symmetric.  Thus the critical point can be viewed as the phase transition point, beyond which the maximally-symmetric spacetimes are not the solutions of the theory.

     We arrived at the above critical point by studying the linearized equations of the EGB theory. It happens that at the critical point the theory
     also admits only one (A)dS vacuum. In \cite{Crisostomo:2000bb}, it was shown that there exist such critical points where only a single (A)dS vacuum was admitted
     in general Lovelock gravities. We may expect that the corresponding theories also have no graviton at the linearized level.

\subsection{Quadratic-order equations at the critical point}

At the critical point, the linearized equations of motion are automatically satisfied, it is thus necessary to study the equations of motion at the quadratic order.  It follows from (\ref{pert}) that we have
\be
g^{\mu\nu}=\bar{g}^{\mu\nu}-h^{\mu\nu}+h^{\mu\sigma}h^{\nu}{}_{\sigma}  + {\cal O}(h^3)\,.
\ee
Up to and including the quadratic order in $h$, we find that the Einstein equations $E_{\mu\nu}=0$ become
\be
\kappa_{\rm eff} \Big({\cal G}^L_{\mu\nu}+\mathcal{G}^Q_{\mu\nu}\Big)-\kappa \Big(\ft12 h_{\mu\nu}R^L-\gamma\big({\cal E}^{(0)}_{\mu\nu}+{\cal E}^{(1)}_{\mu\nu}+{\cal E}^{(2)}_{\mu\nu}\big)\Big) =0 \,.\label{bilinear1}
\ee
Here ${\cal G}_{\mu\nu}^L$ is given by (\ref{gmunuL}) and ${\cal G}_{\mu\nu}^Q = R_{\mu\nu}^Q - \ft12 \bar g_{\mu\nu} R^Q$.  The quantities $R_{\mu\nu}^Q $ and $R^Q$ denote the Ricci tensor and Ricci scalar at the quadratic order of $h$, respectively.  It is clear that the details of these two quantities are irrelevant at the critical point $\kappa_{\rm eff}=0$.  The quantities ${\cal E}_{\mu\nu}^{(i)}$ are given by
\bea
\label{bilinear2} {\cal E}^{(0)}_{\mu\nu}&=&2\Big(R^LR^L_{\mu\nu}-2R^L_{\mu\sigma}\widetilde{R}^{L\,\sigma}_\nu-
2R^L_{\mu\lambda\nu\rho}\widetilde{R}^{L\lambda\rho} +R^L_{\mu\lambda\rho\sigma}\widetilde{R}^{L\lambda\rho\sigma}_\nu \Big)\cr &&
-\ft 12\bar{g}_{\mu\nu}\Big(R^{L}R^{L}-4R^L_{\lambda\rho}\widetilde{R}^{L\lambda\rho} +R^L_{\lambda\rho\sigma\tau}\widetilde{R}^{L\lambda\rho\sigma\tau} \Big) \,,\cr {\cal E}^{(1)}_{\mu\nu}&=&\fft{8(D-3)\Lambda}{(D-1)(D-2)}\Big(R^L_{\mu\lambda\nu\rho}h^{\lambda\rho}
+2R^L_{\sigma(\mu}h^{\sigma}_{\nu)}-\bar{g}_{\mu\nu} R^L_{\lambda\rho}h^{\lambda\rho}-\ft14(D-2) h_{\mu\nu}R^L \Big)\,,\cr {\cal E}^{(2)}_{\mu\nu}&=&\fft{8\Lambda^2}{(D-1)^2(D-2)^2}\Big(\bar{g}_{\mu\nu}h^2+
(D^2-5D+5)\big(\bar{g}_{\mu\nu}h_{\lambda\rho} h^{\lambda\rho}-
2h_{\mu\sigma}h^{\sigma}_\nu \big)\Big)\,.
\eea
Here the curvature tensors with tildes are defined by raising the indices from $R^L_{\mu\nu}$ and $R^L_{\mu\nu\rho\sigma}$ with the background metric $\bar g_{\mu\nu}$.  All the untilded tensor or scalar quantities with the superscript $L$ are given in (\ref{RL1}) and (\ref{RL2}).  Thus we see that the superscript of $i$ in ${\cal E}^{(i)}_{\mu\nu}$ denotes the order of the bare $h$.  It follows that the first term in the second bracket of (\ref{bilinear1}) is similar to  ${\cal E}^{(1)}_{\mu\nu}$.  We separate it out so that we can see clearly that it comes from the Einstein-Hilbert term rather than from the Gauss-bonnet term.  At the critical point (\ref{gammalambda}), only the second bracket in (\ref{bilinear1}) survives.

\subsection{Solutions at the critical point}

The general ``spherically-symmetric'' ansatz can be parameterized as
\be
ds^2 = - h dt^2 + \fft{dr^2}{f} + r^2 d\Omega_{D-2,k}^2\,, \label{sphansatz}
\ee
where $d\Omega_{D-2,k}^2$ with $k=0\,,\pm 1$ denotes the metric of the $(D-2)$-dimensional maximally-symmetric space with $R_{ij} = (D-3)k\, \delta_{ij}$.  (Note that in this paper, we adopt the loose terminology ``spherically-symmetric'' to denote solutions for all $k=1,0,-1$, for the lacking of a simple terminology for general topologies.)
The Schwarzschild-like solutions for the general EGB theory was obtained in \cite{Boulware:1985wk,Cai:2001dz}.  They become degenerate at the critical point.  We find that at the critical point, there are two types of  solutions
\bea
\hbox{type 1}: && h=f=g^2r^2+k-\fft{\mu}{r^{\fft{D-5}{2}}}\,,\qquad \Lambda=-\ft12 (D-1)(D-2) g^2\,,\label{type1}\\
\hbox{type 2}: && f=g^2 r^2 + k\,,\qquad \hbox{and $h$ is an arbitrary function of $r$}\,.
\label{type2}
\eea
We first examine the type-1 solution, which describes a black hole, with the outer horizon located at the largest $r_0$ for which $f(r_0)=0$. This solution was first constructed in \cite{Crisostomo:2000bb,Aros:2000ij}, where Lovelock gravities with single AdS vacuum were studied. The temperature and the entropy can be determined by the standard technique, given by
\be
T=\fft{(D-1)g^2r_0^2+k(D-5)}{8\pi r_0}\,,\qquad S=\ft 14\kappa \omega r_0^{D-2}\Big(1+\ft{(D-2)k}{(D-4)g^2r_0^2}\Big) \,,
\ee
where $\omega$ is the volume of the metric $d\Omega_{D-2,k}^2$. One may determine the mass of the black hole by the completion of the first law of black hole thermodynamics. However, the black hole solution (\ref{type1}) has the unusual falloff, rather than the $1/r^{D-3}$ falloff that is typical of the condensation of the graviton modes.  To derive the first law, we apply the Wald formalism.  The explicit expressions of the Wald formalism \cite{wald1,wald2} for the spherically-symmetric solutions in gravity extended with quadratic curvature invariants were obtained in \cite{fanlu}. It is  given by
\be
\delta {\cal H}=\fft{\omega\, \kappa }{16\pi} r^{D-2} \sqrt{\fft{h}{f}} \Big(
-\fft{D-2}{r} + \fft{2(D-2)(D-3)(D-4)\gamma (f-k)}{r^3}\Big)\delta f\,.
\ee
It is easy to verify that evaluating the above on the horizon yields $\delta {\cal H}_+=T\delta S$; whilst evaluating it at the asymptotic infinity gives
\be \delta H_{\infty}=\fft{\kappa\omega(D-2)}{16\pi g^2}\mu \delta \mu \equiv\delta M\,,\ee
implying that the black hole mass is
\be
M=\fft{\kappa\omega(D-2)}{32\pi g^2}\mu^2 \,.
\ee
The quadratic dependence of the mass on the constant $\mu$ is a consequence that there are no linearized equations of motion in this theory. It follows that the first law of black hole thermodynamics reads $dM=T dS$.  One may also treat the effective cosmological constant $\Lambda$ as a  thermodynamical pressure \cite{Cvetic:2010jb,Kastor:2009wy}, and then the first law becomes $dM=TdS + V dP$, where
\be
P=-\fft{\Lambda}{8\pi}\,,\qquad V=\fft{\kappa \omega r_0^{D-5}}{2(D-1)(D-4) g^4}
\big(k^2 - (D-1) k g^2 r_0^2 -(D-4) g^4 r_0^4\big)\,.
\ee
The Smarr relation becomes
\be
M=\fft{D-2}{D-3} T\, S - \fft{2}{D-3} V P\,.
\ee
When $k=0$, the solution becomes an AdS planar black hole. In this case,an extra scaling symmetry emerges and leads to an additional Smarr relation that is independent of the pressure \cite{Liu:2015tqa}
\be
M=\fft{D-2}{D-1}T S\,. \label{smarr1}
\ee
It is also worth noting that the solution has a space-like curvature singularity at the origin $r=0$.  The singularity is milder than the usual Schwarzschild-like black hole.  We consider $D=5$ as an example, in which case, $g_{tt}$ is non-divergent at $r=0$, and the Riemann tensor squared,
\be
R_4\equiv R^{\mu\nu\sigma\rho}R_{\mu\nu\sigma\rho}\,,\label{riemsq}
\ee
has the $1/r^4$ divergence rather than the $1/r^8$ divergence as $r\to 0$. The free energy of the black hole $F=M-TS$ is given by
\be
F=\fft{\kappa \omega r_0^{D-5}}{32\pi (D-4) g^2} \big( (D-2) k^2 - 6g^2 r_0^2 k -
(D-4) g^4 r_0^4\big)\,.
\ee
Thus for $k=1$, there is also a Hawking-Page-type phase transition \cite{Hawking:1982dh}.  The minimum temperature for the black holes is
\be
T_{\rm min}=\fft{\sqrt{(D-1)(D-5)}\,g}{4\pi}\,,
\ee
under which only the thermal vacuum can exist. For any given temperature above $T_{\rm min}$, there can exist both the thermal AdS vacuum and the black holes of both large and small radii.  There exists a phase transition temperature
\be
T_{\rm phs.}=g\sqrt{\fft{(D-3)(D^2-6D+17)^{\fft32} +
D^4 - 12 D^3 + 34 D^2 + 12 D - 107}{32(D-4)(D-2)\pi^2}}\,,
\ee
above which the black hole with the larger radius develops a negative free energy and hence the thermal vacuum will collapse to form a black hole.  It is worth pointing out however that the Euclidean action is divergent even after subtracting the background values, indicating a possibility of  violating the quantum statistic relation (QSR). (In \cite{Crisostomo:2000bb}, Euclidean action were constructed by introducing boundary counterterms; however there is no covariant expression for such counterterms.) Such violations were reported in analysing the black holes in the Horndeski gravity \cite{Feng:2015oea,Feng:2015wvb}, and in other gravity theories with non-minimally coupled matter \cite{Feng:2015sbw}.

The second type of solutions in (\ref{type2}) are rather unusual, since $h$ can be an arbitrary function of $r$.  For $k=1$ and $h$ being regular for $r\in [0,\infty)$, the solution describe a smooth soliton.  It also allows Lifshitz-type spacetime \cite{Dehghani:2010kd,Dehghani:2010gn,Chen:2016qks} when $h\sim r^{2z}$ with a generic Lifshitz exponent $z$. Since $f$ contains no integration constant, it follows that $\delta f=0$ when evaluating asymptotically, and hence the solution has no mass or any non-trivial conserved charges. These properties imply that the solutions describe the degenerate condensates of the non-dynamical linear modes.

It is remarkable that the two types of solutions (\ref{type1}, \ref{type2}) comprise the most general spherically-symmetric solutions at the critical point, and all of them have the mass $M\ge 0$.  This suggests that the theory at the critical point does not contain nonlinear ghost modes.

Note that the black hole solution is much simpler than the Schwarzschild-like black hole \cite{Boulware:1985wk,Cai:2001dz} in the general EGB theories.  The special critical point was also noticed in \cite{Anabalon:2009kq}, where a rather simple but non-trivial rotating solution was obtained.  The implication of the double zero in the field equations around the AdS in the AdS/CFT correspondence were studied in \cite{Banados:2005rz}.

\section{General quadratically-extended gravity}

In this section, we consider the Einstein gravity extended with the general three quadratic curvature tensor invariants in general dimensions $D$.  The Lagrangian is given by
\be
\mathcal{L}=\kappa\sqrt{-g} \Big((R-2\Lambda_0)+\alpha R^2+\beta\, R_{\mu\nu} R^{\mu\nu}+\gamma\, \big(R^2-4R_{\mu\nu}R^{\mu\nu}+R_{\mu\nu\rho\sigma}R^{\mu\nu\rho\sigma}\big)\Big)\,.
\ee
This theory was well studied in \cite{Deser:2002rt,Deser:2002jk,Gullu:2009vy}.  In this section, we adopt the notation and the linearized formulae in \cite{Deser:2011xc}. The equations of motion are $\kappa\, E_{\mu\nu}=0$, where
\bea
E_{\mu\nu} &=& R_{\mu\nu} - \ft12 g_{\mu\nu} R + \Lambda_0 g_{\mu\nu} +
2\alpha R (R_{\mu\nu} - \ft14 R g_{\mu\nu}) +
(2\alpha + \beta) (g_{\mu\nu} \Box - \nabla_\mu\nabla_\nu) R\cr
&&+\beta \Box( R_{\mu\nu} - \ft12 R g_{\mu\nu}) + 2\beta (R_{\mu\sigma\nu\rho} -
\ft14 g_{\mu\nu} R_{\sigma\rho}) R^{\sigma\rho}\cr
&&+2\gamma \big(R R_{\mu\nu} - 2 R_{\mu\sigma\nu\rho} R^{\sigma\rho} +
R_{\mu\sigma\rho\lambda} R_\nu{}^{\sigma\rho\lambda} - R_{\mu\rho} R_{\nu}{}^\rho\big)\cr
&& - \ft12\gamma g_{\mu\nu} \big(R^2-4R_{\mu\nu}R^{\mu\nu}+R_{\mu\nu\rho\sigma}R^{\mu\nu\rho\sigma} \big)\,.
\eea
There are in general two distinct (A)dS vacua, whose the effective cosmological constants are determined by the quadratic algebraic equation (\ref{vacua}) with $\Delta_0$ now replaced by $\Delta$, given by
\be
\kappa\big(\ft 12(\Lambda-\Lambda_0)+\Delta\Lambda^2\big)=0\,,\qquad \Delta=(D\alpha+\beta)\fft{(D-4)}{(D-2)^2}+\gamma \fft{(D-3)(D-4)}{(D-1)(D-2)} \,.\label{vacua2}
\ee
The formulae (\ref{deltapm}) and (\ref{2in1}) still hold but with $\Delta_0$ being replaced by $\Delta$ and the related discussions are also valid here, except that now $\Delta$ has the $(\alpha,\beta)$ dependence.

\subsection{Conventional critical gravity}

The linearized equations of motion for the metric fluctuations around one of the two (A)dS vacua are \cite{Deser:2011xc}
\be \tilde \kappa_{\rm eff}\, \mathcal{G}^L_{\mu\nu}+\kappa(2\alpha+\beta)\Big(\bar{g}_{\mu\nu}\bar{\square}-\bar{\nabla}_\mu \bar{\nabla}_\nu+\fft{2\Lambda}{D-2}\bar{g}_{\mu\nu} \Big)
R^L+\kappa \beta\Big(\bar{\square} \mathcal{G}^L_{\mu\nu}-\fft{2\Lambda}{D-1}\bar{g}_{\mu\nu}R^L \Big)=0 \,,\label{linear2}
\ee
where
\be
\tilde \kappa_{\rm eff} =\kappa\big(1+4 \tilde \Delta\, \Lambda\big)\,,\qquad
\tilde \Delta =\ft{D\alpha}{D-2}+\ft{\beta}{D-1}+\ft{(D-3)(D-4)\gamma}{(D-1)(D-2)} \,.
\ee
The trace equation turns out to be
\be
\kappa \Big(4\alpha(D-1)+D\beta \Big)\bar{\square}R^L-(D-2) \kappa_{\rm eff}\, R^L=0\,, \label{trace2}
\ee
where $\kappa_{\rm eff}$ has the same definition as (\ref{linear1}), but now with $\Delta$ being given by (\ref{vacua2}). (Note that $\Delta$ and $\tilde \Delta$ become the same when $\alpha=0=\beta$.)  In \cite{Lu:2011zk,Deser:2011xc}, it was proposed to consider
\be
4\alpha(D-1)+D\beta=0 \,.\label{crit1}
\ee
so that the scalar mode becomes non-dynamical.  The equations of motion then implies that $R^L=0$ for a generic $\kappa_{\rm eff}\ne 0$.  It follows from the gauge choice (\ref{gaugechoice}) that $h=0$.  The linearized equations of motion for the transverse and traceless modes now become
\be -\fft{\beta}{2}\Big(\bar{\square}-\fft{4\Lambda}{(D-1)(D-2)}-M^2 \Big) \Big( \bar{\square}-\fft{4\Lambda}{(D-1)(D-2)} \Big)h_{\mu\nu}=0 \,,\ee
where
\be M^2=-\beta^{-1}\Big(\kappa_{\rm eff}+\fft{4\kappa \Lambda\beta}{(D-1)(D-2)} \Big)   \,.\ee
Hence, the theory contains in general one massless and one massive graviton, satisfying respectively
\be \Big( \bar{\square}-\fft{4\Lambda}{(D-1)(D-2)} \Big)h^{(m)}_{\mu\nu}=0\,,\qquad \Big( \bar{\square}-\fft{4\Lambda}{(D-1)(D-2)}-M^2 \Big)h^{(M)}_{\mu\nu}=0 \,.\ee
The absence of the tachyonic mode requires $M^2\geq 0$. When we further requires that $M^2=0$, then the relation
\be \kappa_{\rm eff}+\fft{4\kappa\Lambda\beta}{(D-1)(D-2)}=0 \,,\label{crit3} \ee
defines the critical point at which the theory contains no massive graviton. As the equation on the fluctuation is a fourth order differential equation, there could be other kinds of modes, for example the logarithmic mode like the those in chiral gravity \cite{Li:2008dq,Maloney:2009ck}
and massive gravity \cite{Bergshoeff:2009hq,Bergshoeff:2009aq} in three dimensions.

\subsection{New critical point}

We now consider a new critical condition.  After imposing (\ref{crit1}), instead of imposing (\ref{crit3}), we impose the following condition
\be  \kappa_{\rm eff}=0\,, \label{crit2}
\ee
where $\kappa_{\rm eff}$ is define by (\ref{trace2}).  Explicitly, the constants $\alpha$ and $\beta$ under this critical condition become
\be
\alpha = -\fft{D(D-3)}{(D-1)(D-2)}\gamma \Big(1 - \fft{1}{8\Delta_0\Lambda_0}\Big)\,,\qquad
\beta = \fft{4(D-3)}{D-2}\gamma \Big(1 - \fft{1}{8\Delta_0\Lambda_0}\Big)\,,\label{crit4}
\ee
where $\Delta_0$ is defined in (\ref{vacua}). The new critical condition contains the one discussed in the EGB gravity, and it reduces to that when $\alpha=0=\beta$. As in the case of the EGB gravity, this condition implies that the two (A)dS vacua coalesce into one, with the effective cosmological constant $\Lambda=2\Lambda_0$.  The consequence of this is that the trace equation (\ref{trace2}) becomes automatically satisfied and the trace $h$ of the metric fluctuations becomes non-dynamical at the linear level.
Given the conditions (\ref{crit1}) and (\ref{crit2}) on the parameters and the gauge choice (\ref{gaugechoice}), we find that the linearized equations of motion are
\be
-\ft12\kappa \beta \Big(\bar{\square}-\fft{2D\Lambda}{(D-1)(D-2)} \Big) \Big( \bar{\square}-\fft{4\Lambda}{(D-1)(D-2)} \Big)\tilde h_{\mu\nu}=0\,,\label{grav2}
 \ee
where $\tilde h_{\mu\nu}$ is transverse and traceless.  Thus the theory contains one massless and one massive graviton, satisfying respectively
\be
\Big( \bar{\square}-\fft{4\Lambda}{(D-1)(D-2)} \Big)\tilde h^{(m)}_{\mu\nu}=0\,,\qquad
\Big( \bar{\square}-\fft{2D\Lambda}{(D-1)(D-2)} \Big)\tilde h^{(M)}_{\mu\nu}=0\,.
\ee
The mass square of the massive modes is given by
\be
M^2=\fft{2\Lambda}{D-1} \,.
\ee
It was shown that the generalized Breitenlohner-Freedman bound for a spin-$s$ field in an AdS background is given by  \cite{Chen:2011in,Lu:2011qx}
\be
(M_s^{\rm BF})^2 \ge \fft{(D-2s+5)^2\Lambda}{2(D-1)(D-2)}\,.
\ee
It is clear that $M^2 \ge (M_2^{\rm BF})^2$ for $s=2$ and $D\ge 3$.  It follows that the theory does not contain tachyonic instability, although ghost modes are inevitable. When $\beta=0$ and hence $\alpha=0$, it reduces to the critical point in the EGB theory, in which case even the graviton become non-dynamical at the linearized level.

\subsection{Exact solutions at the new critical point}

Even spherically-symmetric solutions are hard to find in quadratically-extended gravity for generic parameters. When $\gamma=0$, the (A)dS Schwarzschild black hole with an appropriate effective cosmological constant is a solution.  It was recently demonstrated numerically in four dimensions that new black holes beyond the Schwarzschild one exist \cite{Lu:2015cqa,Lu:2015psa}.  This  indicates that a variety of new black holes may exist in generally-extended gravities.  The simplicity of the black hole solution at the critical point of the EGB theory suggests that exact solutions may be easier to construct in the new critical theory.  Indeed for the parameters (\ref{crit4}), we find a new solution under the spherical symmetric ansatz (\ref{sphansatz}) with
\be
f=g^2 r^2 + k\,,\qquad h=(\sqrt{g^2 r^2 + k}+\mu)^2\,.
\ee
where $k$ characterize the topology of the solution, and $g=1/\ell$ is the inverse of the (A)dS radius, related to the effective cosmological constant by
\be
\Lambda=2\Lambda_0=-\ft12(D-1)(D-2) g^2\,.\label{lamg}
\ee
The Riemann curvature squared (\ref{riemsq}) is given by
\be
R_4 = \fft{9g^4}{(\sqrt{g^2r^2 + k}+\mu)^3} \Big(\mu(11 g^2 r^2 + 3\mu^2 + 11k) +
(5g^2r^2 + 9 \mu + 5k) \sqrt{g^2 r^2 + k}\Big)\,.
\ee
It is of interest to note that there is no curvature singularity at $r=0$.
Thus the solution describes a smooth soliton for $k=1$ and $\mu>0$, with the radial coordinate $r$ runs from 0 to asymptotic infinity.  Using the formulae obtained in \cite{fanlu} for the Wald formalism, we find that the mass of the soliton vanishes, or to be precise, $\delta M=0$.

We also obtain (A)dS planar black holes ($k=0$) for some specific $\gamma$:
\be
h=f=g^2 r^2 - \fft{\mu}{r^{D-4}}\,,
\ee
The $(\alpha,\beta,\gamma)$ parameters, satisfying the critical condition (\ref{crit4}), are given by
\be
\{\alpha, \beta, \gamma\}=\fft{1}{\Lambda_0}\{-\fft{D(D-1)}{8(D-2)(D-3)}, \fft{(D-1)^2}{2(D-2)(D-3)},
-\fft{D-1}{4(D-4)(D-3)^2}\}\,.
\ee
The solutions describe the black holes, but also with vanishing mass and entropy, according to the formulae in \cite{fanlu}. These solutions can be viewed as thermalized vacua, similar to those found in the conformal gravity \cite{Lu:2012xu}.

\subsection{A further critical point}

When $\beta=0$ and $\tilde \kappa_{\rm eff}=0$, it follows from (\ref{linear2}) that the linearized equations of motion involve only the trace scalar mode, with no kinetic term for any transverse mode. For non-vanishing $\alpha$ and $\gamma$, we find no exact black hole or soliton solutions.  When $\alpha=0$, the theory reduces to the EGB theory at the critical point.  In $D\le 4$ or $\gamma=0$, the theory reduces to the well-known $f(R)$ gravity with $f=(R- \fft{2D}{D-2}\Lambda)^2$, whose equations of motion reduce to a single scalar equation $R=\fft{2D}{D-2}\Lambda$.  It is of interest to note that the critical case where all linear perturbations have no kinetic terms can only occur when $\alpha=0=\beta$, in other words, in the EGB theory.

\section{Deviation from the critical points}

In the previous sections, we have studied the new critical points of extended gravities where three cases emerge: (1) the whole kinetic terms of all $h_{\mu\nu}$ vanish; (2) that of $h=h^\mu_\mu$ vanishes; (3) that of the transverse $h_{\mu\nu}$ vanishes.  We have obtained a large number of exact static solutions.  In this section, we examine how these solutions change when we deviate from these critical points.  This can help us to determine the physical meaning of the integration constants in the more general setting.

\subsection{Case 1.}

First we consider the parameters
\bea
&&\Lambda_0=-\ft 14(D-1)(D-2)g^2+\fft{(D-2)(D-3)^2(D-1)^3\alpha g^4}{8\kappa(D^2-2D+9)}\,,\quad \beta=-\fft{2(D+1)^2\alpha}{D^2-2D+9}\,,\nn\\
&& \gamma=-\fft{(D-1)^2(3D^2-14D+7)\alpha}{4(D-3)(D-4)(D^2-2D+9)}+\fft{\kappa}{2(D-3)(D-4)g^2} \,.
\label{coupling}
\eea
It reduces to the critical case when $\alpha=0$ and hence $\beta=0$.  In  general dimensions, we find AdS planar black holes ($k=0$), with
\be
h=f=g^2r^2-\fft{\mu}{r^{\fft{D-5}{2}}} \,,\qquad \Lambda=\ft12 (D-1)(D-2) g^2\,.\label{bh2}
\ee
Note that the form of the solution is identical to that in the EGB theory at the critical point.  We now show that the parameter $\mu$ is related to the massive spin-2 modes.

Using the general formula worked out in \cite{fanlu}, we find that the first law of black hole thermodynamics $dM=TdS$ holds, with
\bea
M &=& \fft{(D-2)\omega}{32\pi g^2} \Big(\kappa-\fft{(D-2)(D-3)(D-1)^2}{D^2-2D+9}\alpha g^2  \Big)\mu^2 \,,\cr
T &=& \fft{(D-1)g^2r_0}{8\pi}\,,\qquad S=\ft 14 \omega r_0^{D-2}\Big(\kappa-\fft{(D-2)(D-3)(D-1)^2}{D^2-2D+9}\alpha g^2  \Big) \,.
\eea
To understand the physical meaning of the $\mu$ parameter, let us consider the linearized perturbation around the AdS vacuum, namely,
\be h(r)=g^2r^2+k+h_1(r)\,,\qquad f(r)=g^2r^2+k+f_1(r) \,,\label{larger1}\ee
where $h_1(r)$ and $f_1(r)$ are small perturbation.
For general parameters away from the critical point, $(h_1,h_2)$ are subject to fourth order  differential equations, and the solutions are \cite{fanlu}
\bea
&&h_1=-\fft{m}{r^{D-3}}+\fft{\xi_1}{r^{\fft{D-5-\sigma_1}{2}}}+
\fft{\xi_2}{r^{\fft{D-5+\sigma_1}{2}}}\,,\nn\\
&&f_1=-\fft{m}{r^{D-3}}+\fft{(D-1-\sigma_1)\xi_1}{2(D-1)r^{\fft{D-5-\sigma_1}{2}}}
+\fft{(D-1+\sigma_1)\xi_2}{2(D-1)r^{\fft{D-5+\sigma_1}{2}}}+
\fft{\eta_1}{r^{\fft{D-5-\sigma_2}{2}}}+\fft{\eta_2}{r^{\fft{D-5+\sigma_2}{2}}} \,,
\eea
where the parameter $m$ is associated with the massless graviton mode, and
\bea
&& \sigma_1^2=\kappa\beta^{-1}\Big( 8D(D-1)\alpha+(D-1)(D+7)\beta+8(D-3)(D-4)\gamma-4 g^{-2} \Big)  \,,\cr
&& \sigma_2^2=\fft{\kappa}{4(D-1)\alpha+D\beta}\Big(4(D-2) g^{-2}-(D-1)\big(4(D^2-6D-1)\alpha-(D^2-9D+32)\beta\big)\cr
&&-8(D-2)(D-3)(D-4)\gamma \Big) \,.
\eea
It follows from the falloff behavior that ($\xi_1\,,\xi_2$) and ($\eta_1\,,\eta_2$) are associated with the massive spin-2 mode and massive scalar
mode respectively. For the parameters (\ref{coupling}), we find that $\sigma_1=\sigma_2=0$. The linearized solutions now become
\be\label{genelinear} h_1=-\fft{m}{r^{D-3}}+\fft{\xi_1+\xi_2\log{r}}{r^{\fft{D-5}{2}}}\,,\qquad
f_1=-\fft{m}{r^{D-3}}+\fft{\rho_1+\rho_2\log{r}}{r^{\fft{D-5}{2}}} \,,\ee
where the massive spin-2 mode and massive scalar mode coincide in the metric function $f$ and we collectively denotes them by $(\rho_1\,,\rho_2)$. Since for $\sigma_1=0=\sigma_2$, we have $h_1\ne f_1$ if the solution involves the massive spin-2 but not the scalar modes.  Thus the exact solution we obtained involves both the massive spin-2 and the scalar modes, but not the massless graviton mode.

In the $D=5$ dimension, the solution (\ref{bh2}) is also valid for the spherical and hyperbolic topologies. The solution becomes
\bea
&&ds^2=-fdt^2+\fft{dr^2}{f}+r^2d\Omega_{3,k}^2 \,,\qquad f=g^2 r^2+k-\mu\,,\nn\\
&&\beta=-3\alpha\,,\qquad \gamma=-\alpha+\fft{\kappa}{4g^2}\,,\qquad \Lambda_0=-3g^2+\fft{4\alpha g^2}{\kappa}.
\label{bh3}\eea
The mass, the temperature and the entropy can be calculated using the formulae in \cite{fanlu}, and we find
\bea
 M &=&\fft{3\omega k \alpha }{4\pi}\mu+\fft{3\omega (\kappa-4\alpha g^2)}{32\pi g^2}\mu^2\,,\cr
 T &=& \fft{g^2 r_0}{2\pi}\,,\qquad S=\fft{\omega r_0^3}{4}\Big(\kappa-4\alpha g^2+\fft{3\kappa k}{g^2r_0^2} \Big) \,.
\eea
It is easy to verify that first law $dM=TdS$ holds.  It is of interest to note that for $k\ne0$, the mass formula involves the linear as well as the quadratic term in $\mu$.  If we perform small perturbation around the AdS vacuum, we find that at the linearized order ($k\neq 0$)
\be
h_1=f_1=-\mu-\fft{m}{r^2} \,.\label{larger2}
\ee
This implies that the scalar mode has no independent parameter in this case.

\subsection{Case 2}

In this subsection, we consider the deviation from the critical point discussed in section 3.
We consider the parameters
\bea
\Lambda_0 &=& -\ft14 (D-1)(D+\epsilon-2) g^2\,,\qquad
\alpha = -\fft{(D-1)(\epsilon^2 + (D-3)\epsilon - D)}{8(\epsilon-1) (D-2)(D+\epsilon-3)\Lambda_0}\,,\cr
\beta &=& \fft{(D-1)(\epsilon-2) (D+\epsilon-1)}{4(\epsilon-1)(D-2) (D+\epsilon-3)\Lambda_0}\,,\cr
\gamma&=&\fft{(\epsilon^2-\epsilon-2) (D-1) (D+\epsilon-2)}{8(D-2)(D-3)(D-4)(D+\epsilon-3)\Lambda_0}\,.
\eea
We obtain planar AdS black holes with
\be
f=g^2r^2-\fft{\mu}{r^{D+\epsilon-4}}\,.
\ee
When $\epsilon=0$, the parameters satisfy the critical conditions (\ref{crit1}) and (\ref{crit2}), and the corresponding solution was obtained in section 3.3.
Note that the solution with $\epsilon=0$ was obtained in section 3.3.  The solution with $\epsilon=-\ft12 (D-3)$ was obtained in section 4.1.  When $\epsilon=-1$,
the solutions are also valid for spherical/hyperbolic topologies, namely
\bea
&&
ds^2=-fdt^2+\fft{dr^2}{f}+r^2d\Omega_{D-2,k}^2\,,\qquad f=g^2r^2+k-\fft{\mu}{r^{D-5}} \,,\nn\\
&& \Lambda_0=-\ft 14 (D-1)(D-3)g^2\,,\quad \alpha=\fft{1}{2(D-3)(D-4)g^2}\,,\quad \beta=-3\alpha\,,\quad \gamma=0\,,
\eea
For all these solutions, we find that the entropy and the mass vanish, with non-vanishing temperature
\be
T=\fft{(D+\epsilon-2) g^2 r_0}{4\pi} \,.
\ee

\section{Conclusions}

It is well known that for appropriate range of the coupling constants, the EGB gravity admits two (A)dS vacua, and the linearized perturbations are the massless gravitons.  We find that in one (A)dS vacuum the graviton has positive kinetic energy, whilst in the other it has negative kinetic energy and hence is ghost-like.  There exists a critical point of the coupling constants for which the two (A)dS spacetimes coalesce and the linearized equations of motion become automatically satisfied.  The linear perturbation of the vacuum hence loose its interpretation as a graviton, and the EGB theory at the critical point describes a gravity theory without graviton.

We then derived the perturbative equations of motion at the quadratic order.  We also constructed the most general static solutions with spherical/toric/hyperbolic isometries.  Using the Wald formalism, we demonstrated that these solutions all had non-negative energies, indicating that the theory may not have nonlinear ghost excitations.  One of the solution describes a previously-known black hole with unusual asymptotic falloffs.  We adopted Wald formalism to derive its the mass, entropy and temperature and hence the free energy.  We found that the first law of black hole thermodynamics holds and furthermore there is also Hawking-Page type of phase transition.

We then considered more general theories involving up to quadratic curvature invariants. We found the critical points in the parameter space at which the linearized equation for the scalar trace mode is automatically satisfied.  Interestingly this allows us to find some exact black hole solutions.  Alternatively, for some other choice of parameters, only the scalar trace mode has a kinetic term whilst the transverse modes do not. The case where all linear perturbations have no kinetic terms occurs only in the EGB theory among the general quadratically-extended gravities. We also considered the theories deviated from these critical points, which enables us to understand better the integration constants such as the mass of the black hole at the critical point in a more general setting.

We expect that these critical points commonly exist in Lovelock gravities or in general higher-derivative gravities.  The lacking of any bilinear kinetic term or two-point function is unusual from the point of view of both classical and quantum field theories and its physical implication is not clear at the moment.  Although the critical point in EGB theory lies outside the causality regions \cite{Hofman:2008ar,deBoer:2009pn,Camanho:2009vw,Buchel:2009sk} of the parameter space, the interesting phenomenon warrants further investigations.

\section*{Acknowledgments}

We are grateful to Zhao-Long Wang for useful discussions. Z.Y.F.~and B.C. are supported in part by NSFC Grants No.~11275010, No.~11335012 and No.~11325522. H.L.~is supported in part by NSFC grants NO. 11175269, NO. 11475024 and NO. 11235003.

\end{document}